\documentclass{osa-article}
\journal{oe}
\articletype{Research Article}

\begin{document}

\title{Temporal dissipative structures in optical Kerr resonators with transient loss fluctuation}

\author{Yuanyuan Chen,\authormark{1} Tuo Liu,\authormark{1} Suwan Sun,\authormark{1} and Hairun Guo\authormark{1,*}}

\address{\authormark{1}Key Laboratory of Specialty Fiber Optics and Optical Access Networks, \\
Joint International Research Laboratory of Specialty Fiber Optics and Advanced Communication, \\
Shanghai Institute for Advanced Communication and Data Science, \\
Shanghai University, Shanghai 200444, China\\
}

\email{\authormark{*}hairun.guo@shu.edu.cn} 


\begin{abstract}
Dissipative structures are the result of spontaneous symmetry breaking in a dynamic open system, which is induced by either the nonlinear effect or loss fluctuations. While optical temporal dissipative solitons in nonlinear Kerr cavities has been widely studied, their operation is limited to the red-detuned regime. Here, we demonstrate an emergent dissipative soliton state in optical nonlinear cavities in the presence of loss fluctuations, which is accessible by self-evolution of the system on resonance.
Based on a modified dissipative and Kerr-nonlinear cavity model, we numerically investigate the effect of the loss modulation on the intracavity field pattern, and in transmission observe a single and bright soliton pulse state at the zero detuning.
The effect of the optical saturable absorption is also numerically investigated, which is recognized as an effective approach to the transient loss fluctuation in the cavity.
The estimated power efficiency of the resonant bright soliton can be higher than that of the conventional dissipative Kerr soliton, which is determined by the loss modulation depth and the pump intensity.
The self-starting soliton state on system's resonance is potentially of wide interest, which physically contribute to insights of the temporal structure formation in dissipative cavities.
On application aspect, it may constitute a way to the generation of ultra-fast soliton pulse trains as well as the generation of soliton micro-combs.
\end{abstract}

\section{Introduction}
Dissipative structures are known as a reproducible steady state of a dynamical open system featuring energy or matter exchange in an environment, which have been widely observed and studied in thermodynamics, chemistry and biology \cite{Nicolis1977Self, haken1984science}, as well as in optics \cite{grelu_dissipative_2012}.
In particular, their observations in dissipative and Kerr-nonlinear optical cavities \cite{wabnitz_suppression_1993, barashenkov_existence_1996, leo_temporal_2010, herr_temporal_2014, godey_stability_2014, brasch_photonic_2016} have revealed a phase-locked state \cite{wen_self-organization_2016,taheri_self-synchronization_2017} among a set of Kerr induced intracavity laser components, in the form of a localized temporal soliton pattern and in spectrum corresponding to an equidistant frequency comb structure \cite{herr_temporal_2014}.
Indeed, temporal dissipative solitons in optical microresonators have constituted a way to fully coherent optical frequency combs, with high-compactness and large repetition frequencies in the microwave range \cite{kippenberg_dissipative_2018}, and open the field of optical soliton micro-comb in integrated nonlinear photonics as well as in chip-scale precise time and frequency measurement.
This has unveiled a wide range of applications such as parallel coherent communications \cite{marin-palomo_microresonator-based_2017, fulop_long-haul_2017, fulop_high-order_2018, corcoran_ultra-dense_2020}, optical ranging \cite{suh_soliton_2018, trocha_ultrafast_2018, wang_long-distance_2020} and parallel LIDAR \cite{riemensberger_massively_2020}, low-noise microwave synthesis \cite{liang_high_2015,liu_photonic_2020,jin_Hertz-linewidth_2021}, astronomical spectrograph calibration \cite{obrzud_microphotonic_2019, suh_searching_2019} and photonic neuromorphic computation \cite{feldmann_parallel_2021, xu_11_2021}.
Fundamentally, the solitons are in a double-balance regime in both
the intensity and the phase panel.
While the cavity dissipation is balanced by the optical parametric gain induced by an external pumping source, the dispersive effect is balanced by the nonlinearity.
Yet, access to the soliton state is non-trivial, as essential energy buildup is required to excite the nonlinear effect as well as the symmetry breaking of the system, which implies operations such as the laser frequency tuning \cite{herr_temporal_2014, guo_universal_2016}, modulated laser side-band tuning \cite{stone_thermal_2018}, or equivalently tuning the cavity resonance by thermal or piezoelectric effects \cite{joshi_thermally_2016, liu_monolithic_2020}.

Several approaches have been reported to ease the operation.
One effective way is to apply a fixed modulation on the pump wave that allows for self-evolution of the system to reach the soliton state, free from the laser tuning \cite{taheri_soliton_2015,cole_kerr-microresonator_2018}.
Indeed, the phase and intensity control on the pump wave is widely recognized as a main approach to the motion control of cavity solitons \cite{erkintalo_phase_2021}, which would introduce a modulated continuous-wave (cw) background serving as lattice traps to bond the soliton \cite{jang_writing_2015, taheri_optical_2017, hendry_spontaneous_2018, wang_addressing_2018}.
Fundamentally, intrinsic cavity parameters including the loss rate and the nonlinear coefficient are equivalently modulated via the pump modulation, such that the dynamic of cavity solitons is controlled both in the intensity and in the phase panel.
Alternatively, as a special form of the pump modulation, pulsed pumping scheme has been proposed, which could promote the pump efficiency in the nature of the parametric seeding \cite{papp_mechanical_2013, obrzud_temporal_2017}, and by this way improve the thermal equilibrium in the system.
Other approaches include using an auxiliary laser to counterbalance the intracavity thermal nonlinearity and trigger the soliton burst \cite{zhou_soliton_2019}, or employing laser self-injection locking to the microresonator \cite{maleki_high_2010, liang_high_2015, pavlov_narrow-linewidth_2018, stern_battery-operated_2018, raja_electrically_2019} which induces spontaneous pulling of the laser frequency, and enables turn-key operation for the soliton micro-comb generation \cite{shen_integrated_2020, voloshin_dynamics_2021} in a monolithic photonic chip with hybrid laser-microresonator integration \cite{briles_hybrid_2021, xiang_laser_2021}.

Nevertheless, in a standard optical dissipative and Kerr-nonlinear cavity, it is suggested that the temporal dissipative soliton state (also called the dissipative Kerr soliton, DKS) is on the red-detuned side of the resonance, where the nonlinear induced symmetry breaking is present \cite{wabnitz_suppression_1993, herr_temporal_2014}.
Instead, stable Turing rolls can be accessed on system's resonance, which are triggered by the effect of modulation instability underlying the pump induced parametric gain, and at low intensity could reach the energy equilibrium in the cavity \cite{godey_stability_2014}.
In addition, optical saturable absorption (SA) could introduce direct loss switching in a system, which has been widely applied in mode locked lasers, resulting in dissipative soliton formation by self-evolution of the system \cite{bao2009Graphene, sun2010graphene, qin_electrically_2020}.
It may also contribute to the structure formation in dissipative spatio-temporal systems with the absorptive nonlinearity \cite{firth_spontaneous_1994, firth_optical_1996, spinelli_spatial_1998}.
\textcolor{black}{
Most recently, there emerges studies of temporal dissipative microresonators embedded with the SA effect, which may bring new insights to cavity dissipative structures \cite{kumagai_saturable_2018,xiao_deterministic_2020,Nakashima_21,suzuki_design_2021}}

\begin{figure}[t!]
    \centering
    \includegraphics[width=1 \linewidth]{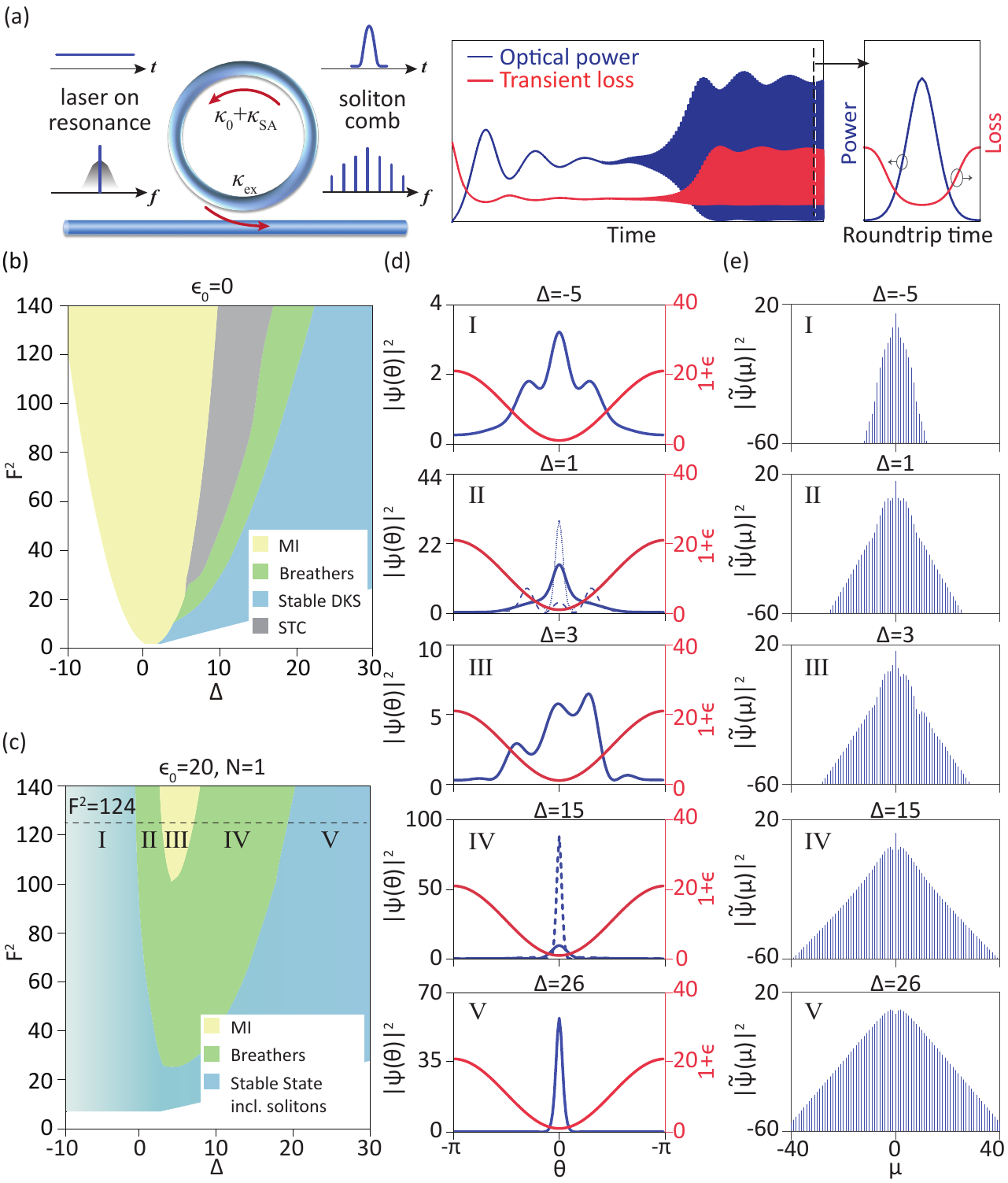}
    \caption{Dissipative structures in microresonators via transient loss. (a) \textcolor{black}{left: Conceptual diagram of an optical microresonator embedded with the SA effect, which could transfer a cw laser on resonance to the soliton microcomb, right: traces of the intracavity optical power (blue curves) and the transient loss (red curves) as a function of time, where the laser features the self-pulsation dynamic assisted by the transient loss.} (b, c) Simulated stability charts of intracavity field patterns as a function of both the laser detuning ${\Delta}$ and the pumping intensity ${F^2}$, in the condition that \textcolor{black}{${{\epsilon}_{0} =0}$} and ${{\epsilon}_{0} =20}$, ${N=1}$. Note: the contrast between the pulse peak to the background (pedestal) is reflected on the lightness of the blue shading area in (c). (d, e) Intracavity temporal profiles and the corresponding spectra of five consecutive states at ${F^2 = 124}$.}
    \label{fig_concept}
\end{figure}

Here, we investigate the soliton regime in dissipative optical cavities in the presence of transient loss modulation effect, and numerically demonstrate a single and bright soliton pulse state that is accessible by natural-evolution of the system on its resonance.
In particular, the estimated efficiency of the resonant bright soliton can be a few times higher than that of the DKS.
We also investigate the the effect of the optical saturable absorption as a mechanism leading to the transient loss modulation, based on a modified Lugiato-Lefever model \cite{lugiato_spatial_1987}, and characterize the soliton state along with other emergent localized temporal field patterns.
Our work may contribute to enrich the dynamics of temporal dissipative solitons particularly in microresonator systems, and may pave a way to direct soliton comb generation in cavities embedded with active low-dimensional materials serving as saturable absorbers.


\section{Numerical model}
The concept of our work is illustrated in Fig. \ref{fig_concept}(a).
\textcolor{black}{In dissipative microresonators with the SA-induced loss dynamic, the cw laser on a cavity resonance could first be self-accumulated to reach the nonlinear threshold, followed by a self-pulsation dynamic in association with the loss profile.
We assume that the response of the SA is sufficiently fast such that the loss has a transient profile within the cavity roundtrip time.
The pulsed laser, once being a stable state of the system, would constitute an optical microcomb in the spectrum.}
We limit ourselves to consider only Kerr nonlinear and second-order dispersive effects in the cavity, with a homogeneous cw driving source.
Such a dissipative system can be described by a modified Lugiato-Lefever equation (LLE) \cite{lugiato_spatial_1987}, with an additional term ${\kappa_v}\left( {t ,\vartheta } \right)$ corresponding to the effect of the loss fluctuation, which is:

\begin{equation}
\frac{\partial A\left( t,\vartheta  \right)}{\partial t}=\left( -\frac{\kappa_v \left( t,\vartheta  \right)}{\text{2}}-\frac{{{\kappa }}}{2}-i{{\delta }_{\omega }}-i\frac{{{D}_{2}}}{2}\frac{{{\partial }^{2}}}{\partial {{\vartheta }^{2}}}+ig{{\left| A\left( t,\vartheta  \right) \right|}^{2}} \right)A\left( t,\vartheta  \right)+\sqrt{\frac{{{\kappa }_{\rm ex}}{{P}_{\text{in}}}}{2\pi \hbar {{\omega }_{0}}}},
\label{lle}
\end{equation}
where ${A\left( t,\vartheta  \right)}$ is the temporal amplitude of the intracavity field, and the integral $\int_{-\pi }^{\pi }{{{\left| A\left( t,\vartheta  \right) \right|}^{2}}d\vartheta }$ indicates the overall number of intracavity photons.
The constant loss rate ${\kappa}$ consists of both the intrinsic loss rate ${\kappa_0}$ in the cavity and the coupling loss rate ${\kappa_{ex}}$, i.e. ${\kappa = \kappa_0 + \kappa_{\rm ex}}$.
The laser cavity detuning, ${\delta}_{\omega}$, is defined as ${{\delta}_{\omega}}={{\omega}_{0}-{\omega}_{p}}$, where ${\omega}_{0}$ is the angular frequency of the pumped central resonance and ${\omega}_p$ is that of the cw pump. ${D}_{2}$ indicates the second order dispersion in the cavity.
The nonlinear coefficient, $g$, indicates the efficiency of the nonlinear induced resonant frequency shift, with respect to the single photon energy ($\hbar{{\omega}_{0}}$).
${P_{\rm in}}$ is the power of the cw pump.

The equation is further normalized to a dimensionless two-parameter model (i.e. in the panel of the laser detuning and the pump intensity), plus a perturbation on the loss, which is:
\begin{equation}
\frac{\partial \Psi \left( \tau ,\varphi  \right)}{\partial \tau }=\left( -{{\epsilon }}\left( \tau ,\varphi  \right)-1-i\Delta -i\frac{{{\partial }^{\text{2}}}}{\partial {{\varphi }^{\text{2}}}}+i{{\left| \Psi \left( \tau ,\varphi  \right) \right|}^{2}} \right)\Psi \left( \tau ,\varphi  \right)+F,
\label{lle_norm}
\end{equation}
with the following transformations:
\begin{equation}
\tau =\frac{\kappa }{2}t,~~
\varphi =\sqrt{\frac{\kappa }{{{D}_{2}}}}\vartheta,~~
\epsilon = \frac{{{\kappa_v}}}{\kappa },~~
\Delta \text{=}\frac{\text{2}{{\delta }_{\omega }}}{\kappa },~~
{{\left| \Psi \left( \tau ,\varphi  \right) \right|}^{2}}=\frac{2g}{\kappa }{{\left| A\left( t,\vartheta  \right) \right|}^{2}},~~
F=\sqrt{\frac{\text{8g}{{\kappa }_{\rm ex}}{{P}_{\text{in}}}}{\text{2}\pi \hbar {{\omega }_{0}}{{\kappa }^{3}}}}.
\end{equation}


In cases of periodically modulated loss factor, we have:
\begin{equation}
{{\epsilon }}\left( \tau ,\varphi  \right)={\epsilon_0}\left(1+\cos \left( N\sqrt{\frac{{D}_{2}}{\kappa}}\varphi +\pi \right)\right),
\label{mod}
\end{equation}
where ${\epsilon_0}$ indicates the modulation depth of the transient loss, $N$ is the number of modulation periods per round trip.
Compared with the pump modulation scheme \cite{erkintalo_phase_2021}, this imposed loss modulation term represents a separated effect in forming intensity traps in the cavity to impact the dynamics of localized field patterns.

It's worth noting that such a modulated loss factor is artificially introduced to the conventional LLE model and is independent on the intracavity field pattern.
In contrast, a self-initiated loss fluctuation can be introduced to dissipative cavities in the regime of saturable absorption (SA), which is related to the intracavity field pattern \cite{f_x_kurtner_mode-locking_1998}, i.e.:

\begin{equation}
\frac{\partial}{\partial \varphi }{{\epsilon }}\left( \tau ,\varphi  \right)=-\frac{{\epsilon} \left( \tau ,\varphi  \right)-\epsilon_0}{{{\varphi }_{SA}}}-{\epsilon} \left( \tau ,\varphi  \right)\frac{{{\left| \Psi \left( \tau ,\varphi  \right) \right|}^{2}}}{{{E}_{\varphi, {SA}}}},
\label{slow}
\end{equation}
where ${{\varphi }_{SA}}$ represents the recovery time of the saturable absorber, and ${{E}_{\varphi, {SA}}}$ is the saturable energy.
In the regime of fast SA (namely the recovery time is shorter than the intracavity pulse duration, leading to ${\left( {\varphi }_{SA} \cdot {\partial {\epsilon}_{\varphi}} \right) \rightarrow 0}$), the loss factor can be further derived as:
\begin{equation}
{{\epsilon }}\left( \tau ,\varphi  \right)=\frac{\epsilon_0}{1+{{\left| \Psi \left( \tau ,\varphi  \right) \right|}^{2}}{{\varphi }_{SA}}/{{E}_{\varphi, {SA}}}}.
\label{fast}
\end{equation}

By replacing the loss perturbation term (${\epsilon }$) in the modified LLE (Eq. \ref{lle_norm}) with Eq. \ref{mod} or Eq. \ref{fast}, one can perform
numerical simulations of the intracavity field pattern in the dissipative Kerr cavity model with the loss modulation or with the fast SA effect.
Technically, the simulation can be carried out by a numerical solver for ordinary differential equations (ODE), which is based on the Runge-Kutta method together with the split-step Fourier transform.
In contrast, simulations with the slow SA effect require solving Eq. \ref{slow} in each round trip during the circulation of the intracavity field, which can be carried out by the solver for boundary value problems (BVP), with the forced condition $\kappa_v \left( \tau ,0 \right)=\kappa_v \left( \tau ,2{\pi } \right)$.

\begin{figure}[t!]
\centering\includegraphics[width= 1 \linewidth]{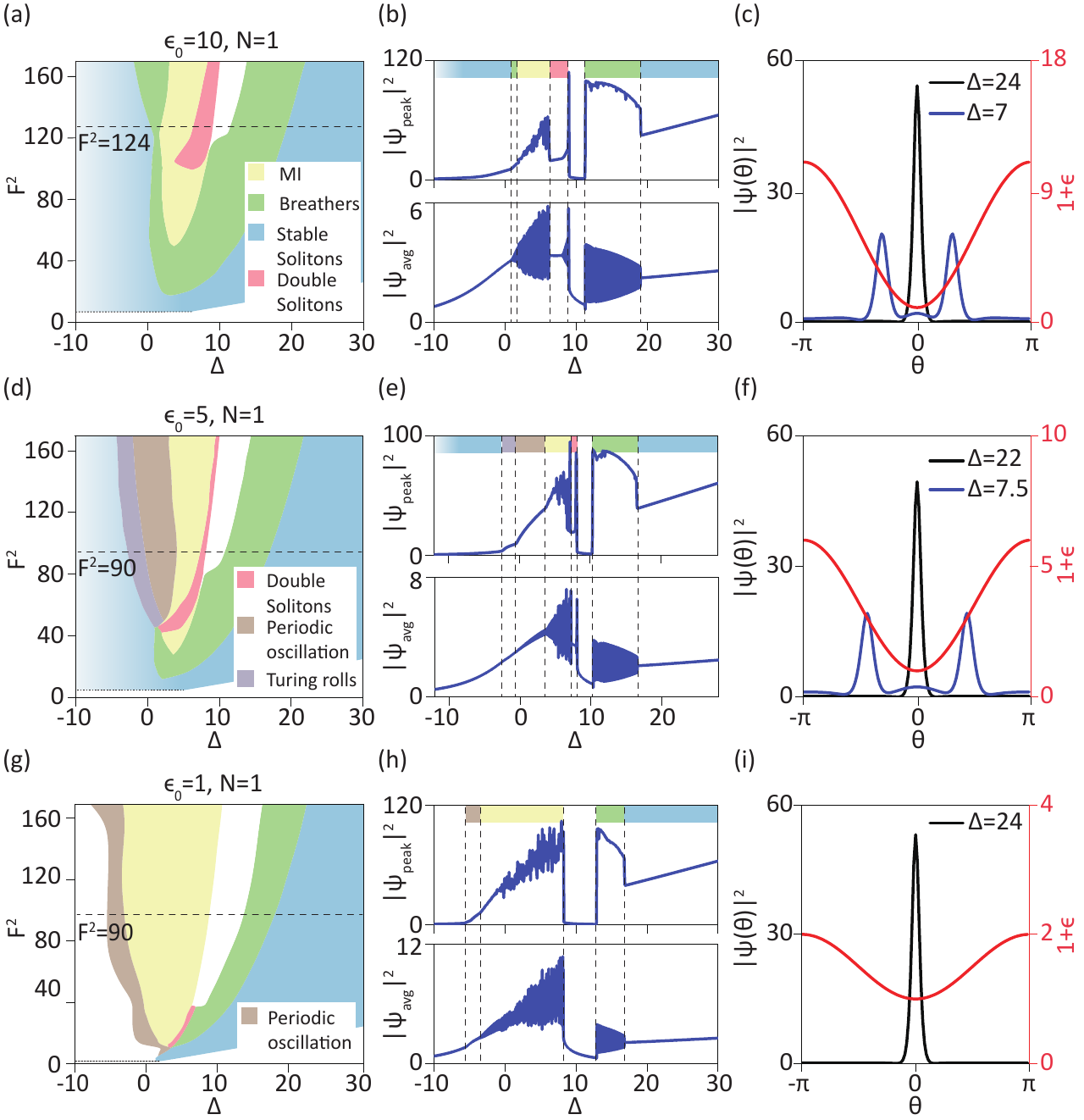}
\caption{Stability charts of dissipative structures.
(a, d, g) Simulated stability charts of intracavity field patterns at three distinct modulation depths of the transient loss fact, i.e. ${{\epsilon}_{0} = 10, ~5, ~{\rm and} ~1}$. (b, e, h) Traces of both the intracavity peak intensity and the averaged intensity selected form each stability chart, as a function of the detuning and at the pump intensity $F^2 = 124$ (b), $F^2 = 90$ (e), and $F^2 = 90$ (h), respectively, in which clear boundaries for distinct dissipative structures can be recognized. (c, f, i) Temporal profiles of the intracavity field pattern at selected detuning values (black and blue curves), together with the modulated loss profile (red curves). }
\label{chart}
\end{figure}

\begin{figure}[t!]
\centering
\includegraphics[width=1 \linewidth]{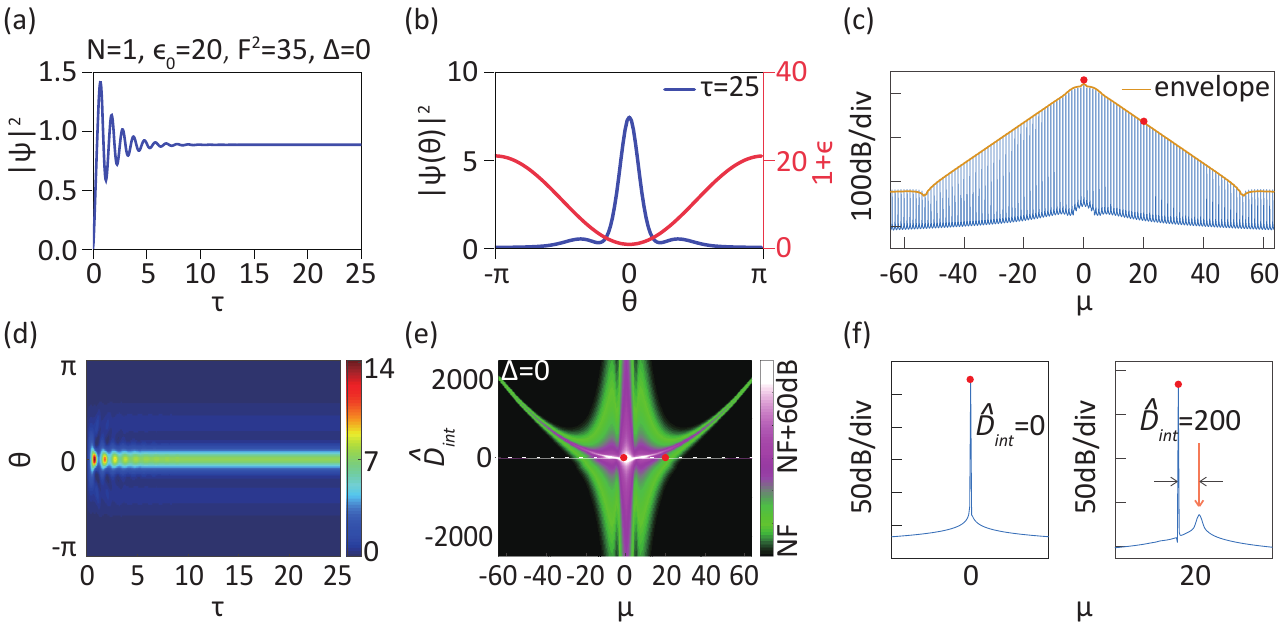}
\caption{Self-starting single-pulse soliton state at the zero detuning.
(a, d) Evolution of both the averaged intensity and the intracavity field pattern over cavity round trips, where ${\Delta = 0}$. (b) The stabilized single-pulse profile after a few round trips. (c) Simulated frequency comb spectrum in a resolution $\frac{1}{2^{14}}$ of the repetition frequency. (e) The spectrogram extracted from the comb spectrum, which reveals the spectral distribution of the laser-to-cavity detuning as a function of the mode index. Note: the intensity is in the logarithmic scale and the color data range is 60 dB based on the level of the noise floor (NF). The distribution is equivalently to the integrated (normalized) dispersion profile ${\hat D_{\rm int}} = \frac{1}{2}\mu^2$. (f) Spectra around selected comb modes with corresponding resonances.
}
\label{self}
\end{figure}

\begin{figure}[t!]
\centering
\includegraphics[width=1 \linewidth]{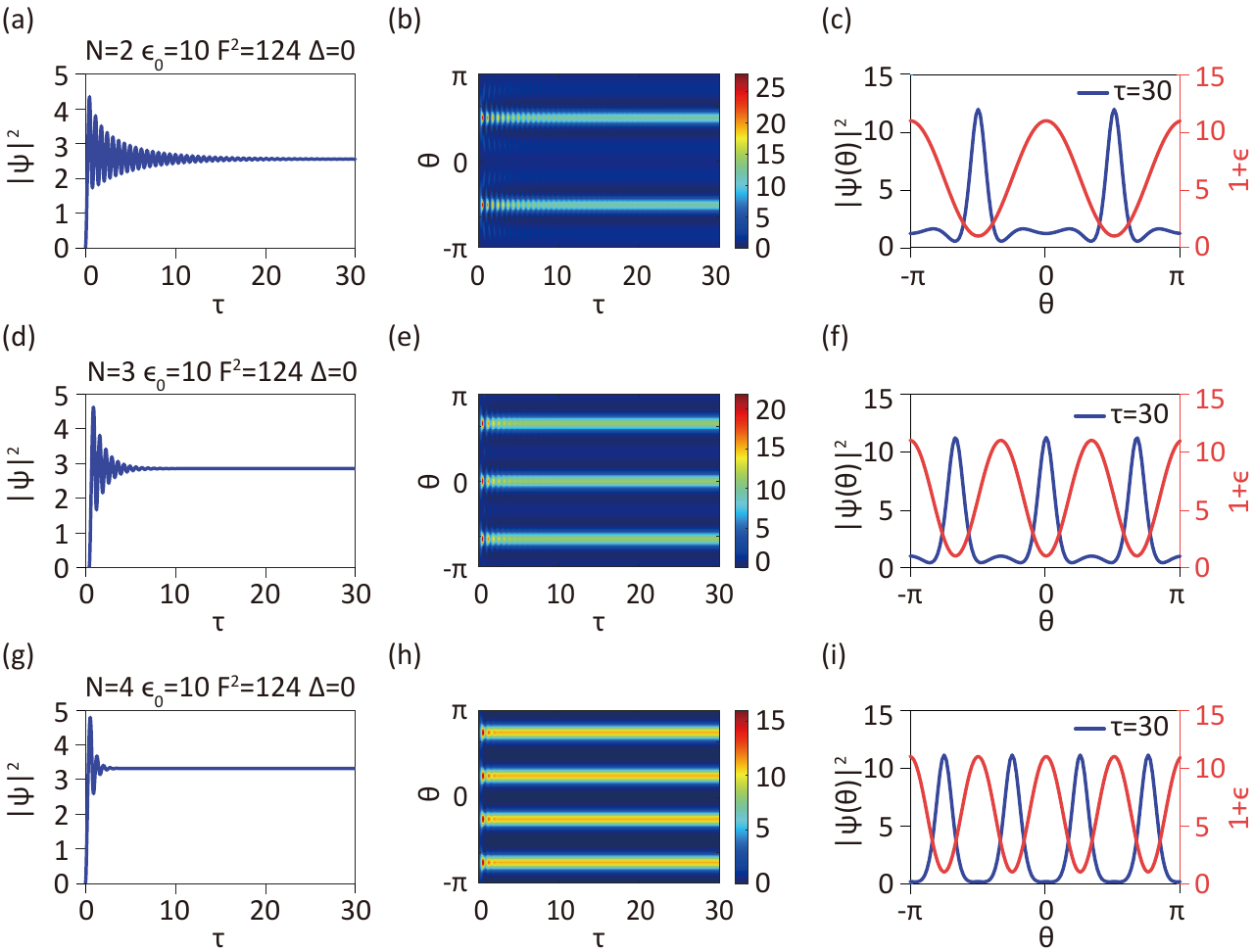}
\caption{
Self-starting multi-pulse soliton state on resonance. (a, d, g) Traces of the averaged intensity of the self-starting soliton state as a function of the time. The loss modulation period is ${N = 2, ~3, ~4}$, and the modulation depth is fixed as $\epsilon_0 = 10$, the detuning is always ${\Delta = 0}$ and the driving intensity is fixed as ${F^2 = 124}$.
(b, e, h) The evolution of the intracavity field pattern, each corresponding to the state on its left.
(c, f, i) The stabilized temporal field pattern of the soliton state (blue curves), together with the modulated loss profiles (red curves).
}
\label{Delta=0}
\end{figure}

\begin{figure}[t!]
\centering
\includegraphics[width=1 \linewidth]{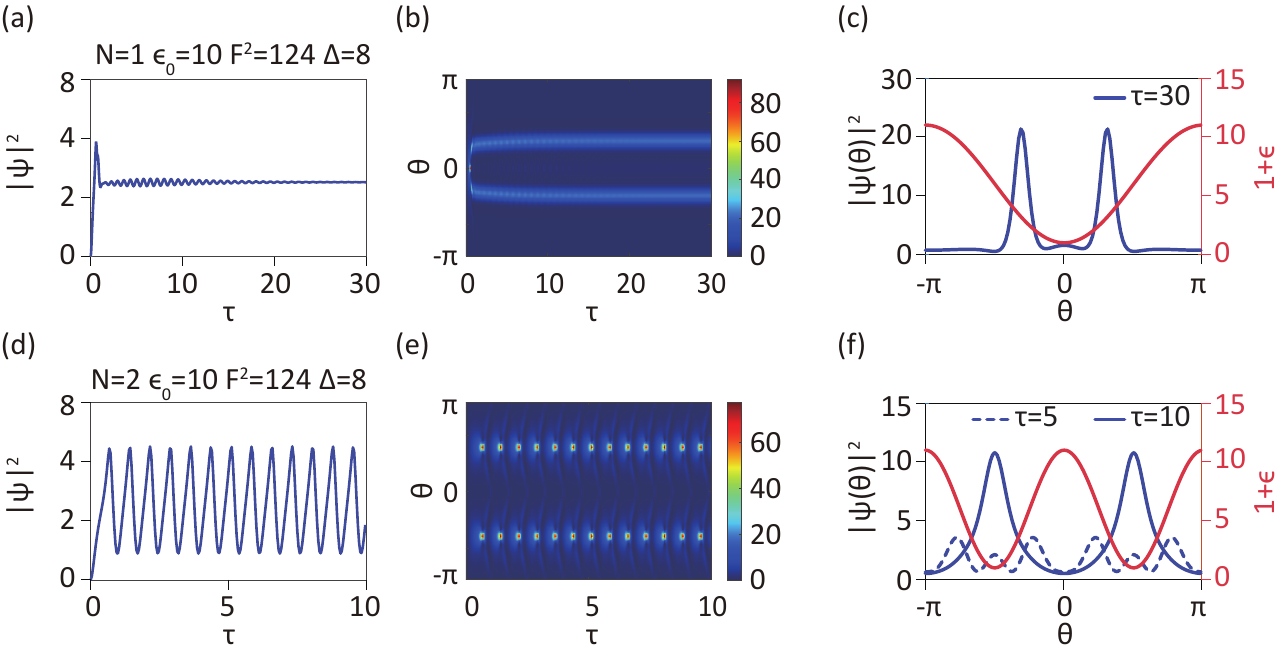}
\caption{
Self-starting dual-pulse state. (a, b) Evolution of both the averaged intensity and the intracavity field pattern over the time, in the condition ${N=1}$ and ${\Delta = 8}$. (c) The stabilized dual-pulse profile (blue curve) together with the loss profile (red curve), at time $\tau = 30$. (d, e, f) A similar investigation to (a, b, c) in the condition ${N=2}$, where the excited dual-pulse state is periodically oscillatory. In detail, the field pattern (f) mastered by one period of the loss modulation is switching between the single-pulse state and the dual-pulse state.
}
\label{Delta=8}
\end{figure}

\begin{figure}[t!]
\centering
\includegraphics[width=1 \linewidth]{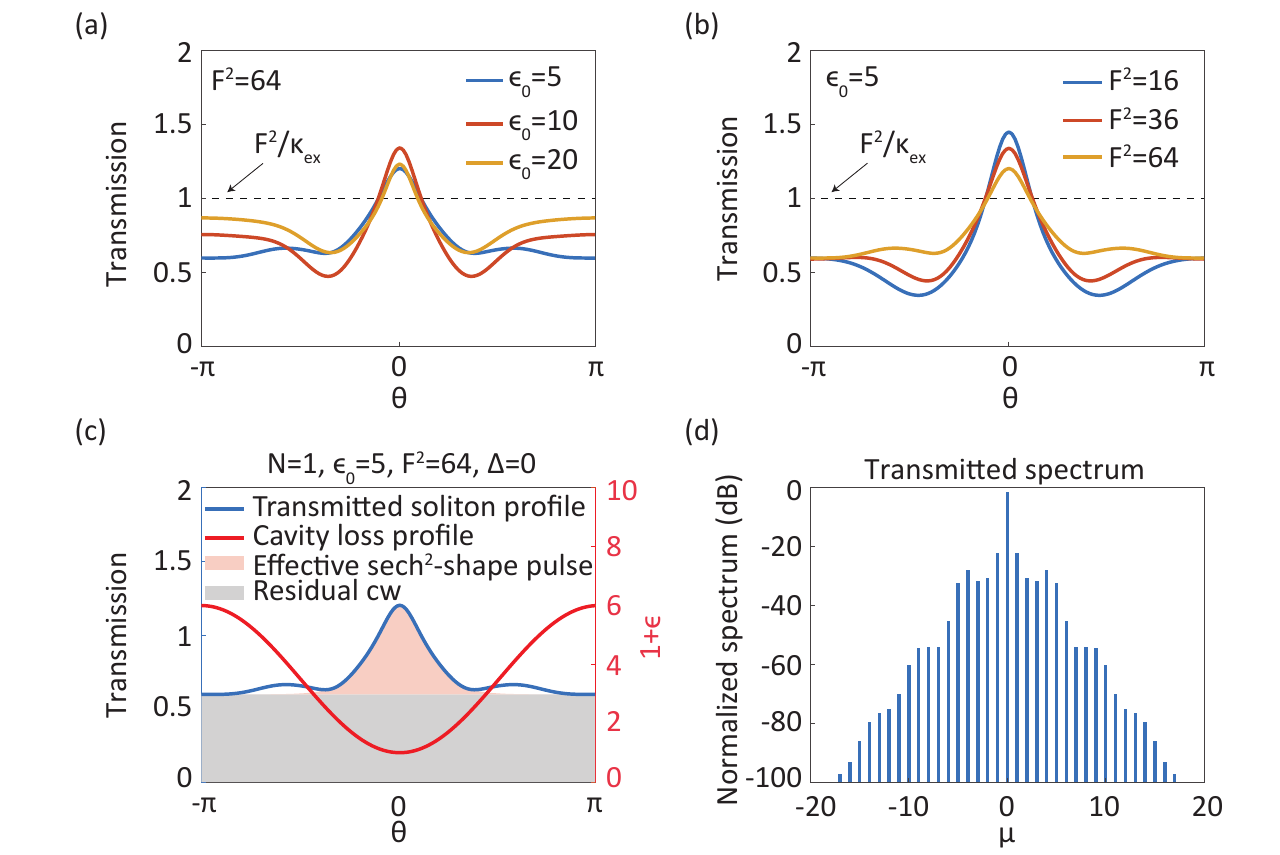}
\caption{
Single-pulse resonant bright soliton state in the transmission. (a) Normalized transmitted soliton pulse profiles with different modulation depths in the loss profile, at the zero detuning and at the same pump intensity $F^2 = 64$. (b) Transmitted soliton pulse profiles and at different driving intensities, with $\epsilon_0 = 5$. (c) A $sech^2$-shape pulse profile (red-shading area), together with the residual cw wave (gray-shading area), is extracted from the transmitted pulse profile (blue curve). (d) Normalized transmitted spectra of the RBS state in (c). Note: the normalization is with respective is with respect to the external driving intensity ($F^2/\kappa_{\rm ex}$, where $\kappa_{\rm ex} = 0.72$ corresponding to the effective critical coupling regime of the loss-modulated cavity system, cf. table \ref{efficiency}).
}
\label{efficiency}
\end{figure}

\begin{table}[ht]
\centering
\caption{
Power efficiency regarding both the RBS and the DKS. The bold numbers indicate that the efficiency of the RBS is higher than the DKS under the same driving intensity.
}
\begin{tabular}{ccccccc}
    \hline
    $\epsilon_{0}$ & $\Delta$ & $F^2$ & ${\kappa_{\rm eff}}$ & $\eta=\kappa_{\rm ex}/\kappa$ & $\Gamma_{\rm RBS}$ & $\Gamma_{\rm DKS}$\\
    \hline
    20 & 0 & 64 & 1.44 & 0.72 & \textbf{0.0681} & / \\
    20 & 0 & 36 & 1.44 & 0.72 & 0.0521 & / \\
    20 & 0 & 16 & 1.44 & 0.72 & 0.0555 & / \\
    10 & 0 & 64 & 1.44 & 0.72 & \textbf{0.1164} & / \\
    10 & 0 & 36 & 1.44 & 0.72 & \textbf{0.1109} & / \\
    10 & 0 & 16 & 1.44 & 0.72 & 0.0794 & / \\
    5 & 0 & 64 & 1.44 & 0.72 & \textbf{0.0908} & / \\
    5 & 0 & 36 & 1.44 & 0.72 & \textbf{0.1347} & / \\
    5 & 0 & 16 & 1.44 & 0.72 & \textbf{0.1541} & / \\
    / & $\Delta_{max}\approx79$ & 64 & 1 & 0.5 & / & $2\eta^2/F=0.0625$ \\
    / & $\Delta_{max}\approx44$ & 36 & 1 & 0.5 & / & $2\eta^2/F=0.0839$ \\
    / & $\Delta_{max}\approx20$ & 16 & 1 & 0.5 & / & $2\eta^2/F=0.1250$ \\
    \hline
   \end{tabular}
   \label{effi_tab}
\end{table}

\section{Results and discussion}
\subsection{Temporal dissipative structures via loss modulation}
We first carried out simulations based on the LLE model with and without the transient loss modulation.
In the absence of the loss modulation, i.e. ${\epsilon_0 = 0}$, the model is degraded to the standard LLE.
The stability chart of the system is simulated, see Fig. \ref{fig_concept}(b), which has a good agreement with previous reported data \cite{godey_stability_2014, karpov_dynamics_2019}.
In contrast, in the condition ${\epsilon_0 = 20}$ and ${N = 1}$, the stability chart is largely changed, see Fig. \ref{fig_concept}(c).
Significantly, the region corresponding to chaotic intracavity fields (i.e. the MI state) is largely shrunk, while that of breathers \cite{akhmediev_universal_2011,leo_dynamics_2013, bao_observation_2016, yu_breather_2017, lucas_breathing_2017} and stable pulse states are increased.
Of particular interest is the fact that the stable pulse state that could exist over a wide range of the detuning (${\Delta}$) and even on the blue-detuned side of the resonance (${\Delta < 0}$).
The pulse peak is located on the minimum of the intracavity loss profile.
Yet, there is a pedestal in the pulse profile, which is distinguishable when the state is in the blue-detuned regime.
In the red-detuned regime, the pedestal can be suppressed such that the pulse peak to the background is increased with an increase of the detuning, which is in accordance to the observed dynamic of the DKS \cite{lucas_detuning-dependent_2017}.

We showcased a typical set of five consecutive intracavity field patterns from blue detuned side to the red-detuned side, at a constant pump intensity, see Fig. \ref{fig_concept}(c, d).
As a result, the stable single-pulse state on the blue-detuned side exhibits a narrower comb spectral bandwidth compared with that on the red-detuned side, and both temporal field patterns are characterized to be the Fourier limited pulses with respect to their spectra.
In the breathing state, the averaged comb spectrum features a triangular envelope (on a logarithmic scale) as its signature, which is similar to previous experimental observations in dissipative and nonlinear microresonators \cite{lucas_breathing_2017}.

We also investigated the stability chart of the intracavity field patterns at different loss modulation depths, see Fig. \ref{chart}.
Together with the chart in Fig. \ref{fig_concept}(b), an overall evolution of the distribution of the intracavity field patterns is revealed.
Given the fact that the overall loss in the system is reduced when the loss modulation depth is decreased, the stability chart is more developed at the high pump intensity side, and overall the intracavity field patterns are enriched.
This include the emergence of a second stable soliton region (cf. the red colored region in Fig. \ref{chart}(a)) and the void region, and the region corresponding to Turing rolls \cite{godey_stability_2014} (cf. the brown and the purple regions in Fig. \ref{chart}(d, g)) by further decreasing the modulation depth (in this trend the system is close to the standard dissipative Kerr nonlinear cavity).

The rich dissipative structures via the loss modulation is determined by a large number of LLE simulations.
In each simulation, the stabilized single pulse state in the red-detuned regime is first excited (which is probed by the DKS pulse profile \cite{herr_temporal_2014} and is stabilized by self-evolution of the system over a number of roundtrips).
Afterwards, \textcolor{black}{while the pump power remains unchanged,} the detuning is backward tuned to sweep over the system's resonance \cite{guo_universal_2016}, in which the evolution of the intracavity state can be observed.
We traced both the peak intensity and the averaged intensity of the intracavity field pattern, which indicate clear boundaries of different dissipative structures in the system.
\textcolor{black}{Each found structure is further verified to be self-consistent by self-evolution of the system without changing the detuning.
The above procedure is repeated for different pump intensities.
Eventually, a stability chart as a function of the pump intensity and the laser detuning is obtained.}

Fundamentally, as the response to the loss modulation, the intracavity field driven by a cw source will form a pulsed pattern, which serves as the preform of the single pulse state.
Therefore, direct coupling from the CW pump to the pulsed soliton state can be implemented.
This process can be resistant to the modulation instability underlying the cw-pump induced parametric gain, indicating an expanded region of the stabilized single-pulse state in the stability charts when the loss modulation depth is increased.

Another emergent state is a dual-pulse state (cf. red shading areas in the stability charts Fig. \ref{chart}(a, d, g)), which is understood to stem from the pedestal of the single-pulse pattern.
Remarkably, the pulses are on the slope of the loss profile with certain delay in between, indicating the existence of an inner bond that resists the entrapping force by the loss modulation.

\subsection{Self-starting single-pulse state on resonance}
In principle, spontaneous symmetry breaking is the key to trigger the formation of dissipative structures, which can be induced by nonlinear effects in the cavity or via loss fluctuations.
The former regime is famous for dissipative microresonator systems where the DKS state exists in the red-detuned regime, while the latter is commonly applied in mode-locked laser systems by means of the saturable absorption.
Here, we aimed to combine a dissipative system with loss fluctuations, and investigate dissipative soliton formation operated on system's resonance (i.e. at the zero detuning ${\Delta = 0}$).
As discussed in the above section, a stable single-pulse state is supported in the system at the zero detuning, while the nonlinear induced bifurcation is NOT presented (the bifurcation only occurs at ${\Delta > \sqrt{3}}$ driven by a homogeneous cw pump \cite{herr_temporal_2014}).
\textcolor{black}{
This indicates that the single-pulse state is the only solution of the system at the zero-detuning region, which can be deterministically accessed by self-evolution of the system, from a noise background and driven by the cw pump.}

Indeed, we observed such a self-starting process regarding the single-pulse state at the zero detuning, see Fig. \ref{self}(a, d), with the loss modulation depth ${\epsilon_0 = 20}$.
The intracavity temporal pulse profile is presented in Fig. \ref{self}(b), in which the single pulse shows a high contrast between its peak to the background (where a small pedestal remains).
Moreover, We simulated a pulse train with ${{2}^{14}}$ cavity round trips, and obtained a noise resolved comb spectrum, \textcolor{black}{i.e. the spectral resolution is ${\frac{1}{2^{14}}}$}, see Fig. \ref{self}(c).
In this way, ``cold'' resonances on the noise background between comb modes can be resolved, and as expected, the lasing modes are found deviated from their resonances, see Fig. \ref{self}(f).
\textcolor{black}{
The comb spectrum can be further transferred to a 2D spectrogram (frequency offset vs. mode index) to highlight detailed background structure around each comb mode, in which the distribution of the laser-cavity detuning as a function of the mode index is revealed, see Fig. \ref{self}(e).}
Remarkably, the distribution is equivalent to the integrated dispersion of the cavity system \cite{chembo_spatiotemporal_2013, herr_temporal_2014} and, in the present case, showing a parabolic profile corresponding to the second order dispersive effect in the cavity (which in the normalized form reads ${\hat D_{\rm int}} = \frac{1}{2}\mu^2$).
This confirms that the on-resonance single-pulse pattern has balanced the cavity dispersive effect with its nonlinear effect, resulting in a localized temporal soliton state.

In the condition of multiple loss modulation periods, i.e. ${N = 2, ~3, ~{\rm and} ~4}$, the on-resonance soliton state is also investigated, see Fig. \ref{Delta=0}.
As a response to the transient loss profile, the soliton state features two-, three-, and four-pulse patterns according to the number ($N$) of the modulation period, in which the pulse peak is always allocated at the minimum of the loss profile.
Similarly, these soliton states are accessible by self-evolution of the system under a homogeneous cw driving force.

The self-starting dynamic of the emergent dual-pulse state is also investigated, see Fig. \ref{Delta=8}.
According to the stability chart, the state exists within a limited range in the red-detuned regime (cf. the red-shading area in Fig. \ref{chart}(a)).
In the condition $N=1$, the state features a self-starting and stabilized dual-pulse pattern (Fig. \ref{Delta=8}(a, b)), while in the condition $N=2$, it turns to two oscillatory single-pulse patterns instead of the dual-pulse patterns (Fig. \ref{Delta=8}(d, e)).
In detail, allocated to each minimum of the loss profile, the oscillatory single-pulse pattern features a transition to the dual-pulse pattern, forth and back.
This transition is understood as when the loss modulation period is shortened (by half), the dual-pulse pattern stemming from the soliton pedestal will feature interactions with the adjacent pattern, and becomes degenerate to the breathing single-pulse pattern that exists within the similar detuning range.

\subsection{Efficiency of single-pulse soliton state on resonance}
We next estimated the power efficiency regarding the single-pulse soliton state on resonance.
In contrast to the conventional DKS, the soliton state shows a bright pulse profile both in the cavity and in the transmission, termed as \emph{resonant bright soliton} (RBS) state, which at certain conditions of the loss modulation depth and the driving intensity may have an outstanding pulse profile with low pedestal, see Fig. \ref{efficiency}(a, b).
Yet, we noticed that the comb spectrum of the RBS dose not have a smooth envelope at its center, see Fig. \ref{efficiency}(d), which is understood as the result of the loss modulation that introduces strong side-bands in the comb spectrum.

\begin{figure}[ht]
\centering
\includegraphics[width=1 \linewidth]{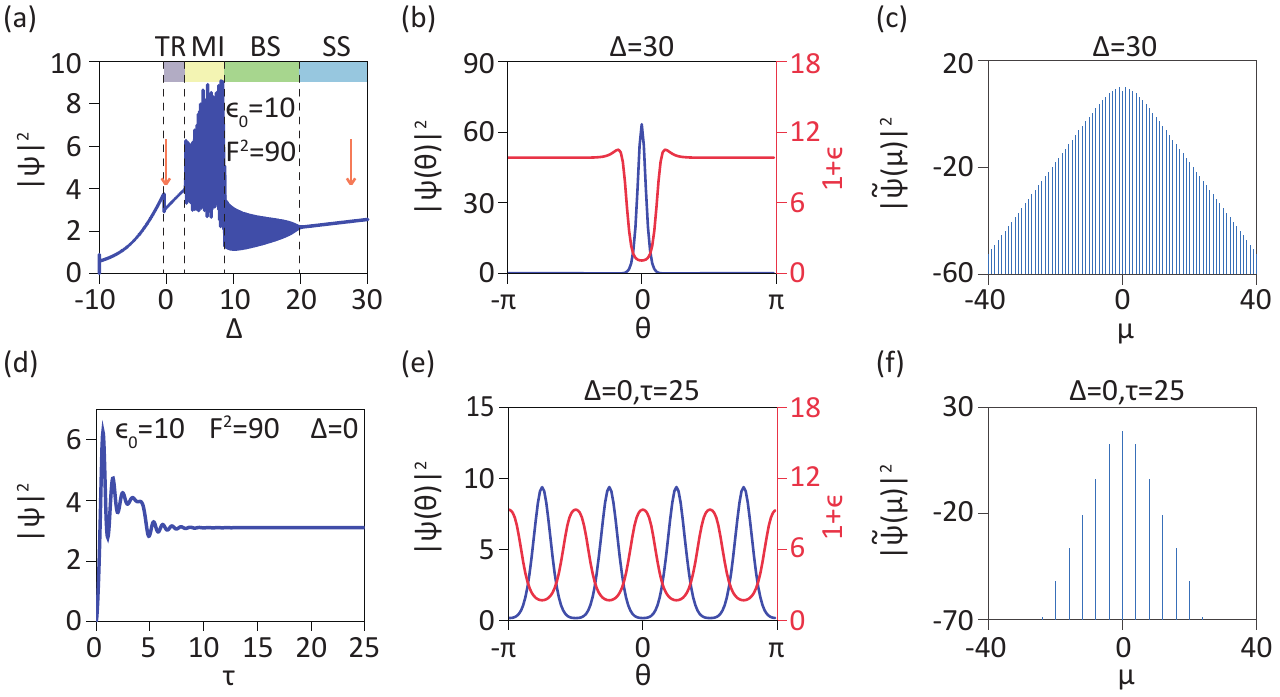}
\caption{Dissipative solitons via fast saturable absorption (SA). (a) The averaged intracavity intensity upon the forward scan of the detuning, in the condition that the modulation depth of the saturable absorption is set as ${\epsilon_0} = 10$. TR: Turing rolls. BS: Breathers. SS: Stable solitons. (b, c) The stabilized temporal soliton pattern along with the intensity-dependent loss profile, and the corresponding comb spectrum, at the detuning $\Delta =30$. (d) Trace of a dissipative structure reached in the system by self-evolution, operated on resonance. (e) The stabilized Turing rolls and the corresponding loss profile upon the self-starting process. (f) The comb spectrum of the Turing rolls. The saturable intensity is ${{I}_{\varphi ,SA}}={{E}_{\varphi ,SA}}/{{\varphi }_{SA}} \approx 0.7$}
\label{fig-fast}
\end{figure}

\begin{figure}[ht]
\centering
\includegraphics[width=1 \linewidth]{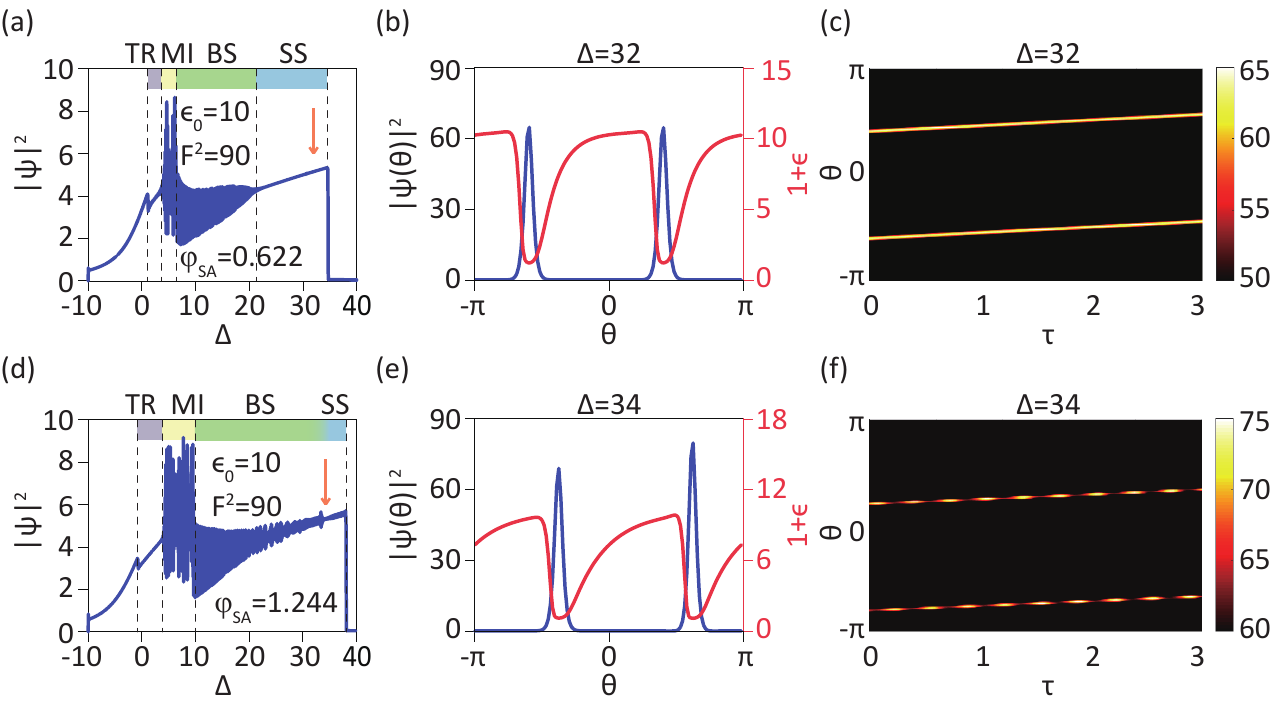}
\caption{Dissipative solitons via slow saturable absorption (SA). (a) The averaged intracavity intensity upon the forward scan of the detuning. The modulation depth of the SA is ${\epsilon_0} = 10$, the recovery time is ${{\varphi }_{SA}} \approx 0.6$ and the saturable energy is ${{E}_{\varphi ,SA}}\approx0.8$. TR: Turing rolls. BS: Breathers. SS: Stable solitons. (b) A two soliton pulse state circulating in the cavity, along with the SA-induced loss profile that is asymmetric with respect to  the soliton profile. (c) Temporal evolution of the soliton pair over time, when the detuning is fixed at $\Delta = 32$. (d, e) Similar to (a, b) in the condition the recovery time is ${{\varphi }_{SA}} \approx 1.2$. (f) Temporal evolution of the soliton pair that features out-of-phase oscillations in the soliton intensity, when the detuning is fixed at $\Delta = 34$. }
\label{fig-slow}
\end{figure}

The soliton power is then estimated by numerically extracting a bright ${\rm sech}^2$-shape pulse profile in the transmitted field pattern, see Fig. \ref{efficiency}(c), \textcolor{black}{and the power efficiency is calculated as the ratio of the soliton power divided by the pump power (i.e. $\Gamma = \frac{P_{\rm soliton}}{P_{\rm pump}}$).}
Detailed values of the efficiency regarding the RBS in different conditions are listed in Table. \ref{effi_tab}.
Significantly, in the presence of the loss modulation, an effective cavity loss rate is first estimated, by modelling the modified LLE (Eq. \ref{lle_norm}) without the dispersive and the nonlinear effects.
Then the external coupling rate ($\kappa_{\rm ex}$) is always set as half of the loss rate ($\kappa_{\rm eff}$), indicating that the system is effectively operated in the critical coupling regime.
This leads to a difference in the ratio $\eta = \frac{\kappa_{\rm ex}}{\kappa}$ when extracting the external pump power as $\frac{F^2}{\eta}$.

As a comparison, the efficiency of the DKS at its maximal detuning is also calculated, which is theoretically derived based on the standard LLE \cite{wabnitz_suppression_1993, bao_nonlinear_2014, herr_temporal_2014, yi_soliton_2015, yi_theory_2016}, and in the normalized two-parameter $(\Delta, F)$ form reads: $\Gamma =\frac{4{{\eta }^{2}}\sqrt{2\Delta }}{\pi {{F}^{2}}}$.
This equation would have good accuracy when the driving force is sufficiently large and supports the DKS at a relatively large detuning $\Delta\gg0$. The theoretical maximal detuning is $\Delta_{max}={\pi^2}{F^2}/8$, at which the efficiency is further derived to be: ${{\Gamma }_{\max }}=2{{\eta }^{2}}/F$.

As a result, we observed that at the same driving source, the RBS can have a higher efficiency than the DKS (cf. values marked in bold in Table. \ref{effi_tab}), though the effect is not dramatic and is dependent on the level of the loss modulation depth.
With an increase in the loss modulation depth, i.e. the overall cavity loss is increased, the efficiency of the RBS can be largely decreased.
In practice, the efficiency is also highly dependent on cavity properties such as the material intrinsic loss, the dispersion, the nonlinearity, and the coupling regime.
Therefore, engineering on cavity parameters is further required to promote the efficiency of both the RBS and the DKS.

\subsection{Temporal dissipative structures via saturable absorption}
In general, the effect of the saturable absorption (SA) is to introduce an intensity-dependent transmittance to a system and is therefore beneficial to realize pulsed structures, which in nature is similar to the effect of the loss modulation.
Indeed, the SA-induced loss fluctuation is simultaneously evolved with the formation of the cavity field pattern.
As such, a positive feedback regime is established in the system, which could initiate the formation of temporal dissipative structures.

To verify this point, we carried out simulations with the SA effect.
The results are presented in Fig. \ref{fig-fast} and Fig. \ref{fig-slow}.
We first performed the forward laser tuning to investigate possible temporal dissipative structures supported in the system, in the condition of a moderate modulation depth in the SA, i.e. ${\epsilon_0 = 10}$.
During the process, multi-pulse RBS on resonance (close to the Turing rolls), MI state, breather state, and stabilized soliton state (close to the DKS) in the red-detuned regime are observed, see Fig. \ref{fig-fast}(a).
As a transient effect, the fast SA-induced loss profile exhibits a decrease along with the soliton pattern and is symmetric regarding the raising and the ending edges of the soliton pulse profile, see Fig. \ref{fig-fast}(b, c).
At the zero detuning, a self-starting and stabilized multi-pulse pattern is observed, see Fig. \ref{fig-fast}(d, e, f).
We noticed that the pattern in its profile is close to the Turing rolls that is usually accessible in the low-detuning region in the standard system \cite{godey_stability_2014}.
While it remains difficult to distinguish whether the state is underlay by the Turing pattern or the RBS, we noticed an elongated existence region of this state compared with the MI state, \textcolor{black}{especially when the recovery time of the SA is increased, see Fig. \ref{fig-slow}(a, d).}
This confirms that the SA-induced loss modulation effect could effectively suppress the chaotic field patterns in the cavity and prefers stabilized dissipative structures.
\textcolor{black}{We note that a similar conclusion has recently been reached in \cite{Nakashima_21}, where by the SA effect the chaotic state is suppressed and a high number of DKS can be accessed.}


Similarly, in the presence of the slow SA effect, both the multi-pulse pattern (at zero detuning) and the DKS-like pattern (in the red-detuned regime) are observed in the system.
In contrast to the fast SA, the slow SA-induced loss profile is asymmetric with respect to the soliton pulse profile, see Fig. \ref{fig-slow}(b, e).
Remarkably, the DKS-like state (in the form of two soliton pulses circulating in the cavity) are not fully stabilized but featuring slow fluctuation in the intensity, see Fig. \ref{fig-slow}(c, f).
This can be attributed to the relaxation process of the slow SA, namely a long recovery time in the loss profile would interact with the neighbouring soliton pulse in the cavity, leading to a coupled soliton pair featuring energy exchange in between.
Indeed, in the condition of a long recovery time in the SA, out-of-phase oscillations in the intensity of the solitons are observed, which is known as the signature for the energy exchange dynamic \cite{guo_intermode_2017}, see Fig. \ref{fig-slow}(f).

\section{Conclusion}
In conclusion, we have investigated rich dissipative structures in an optical cavity system with transient loss fluctuation.
With a direct modulation on the loss factor, an emergent single-pulse and bright soliton state on system's resonance was observed and characterized, which can be accessed by self-evolution of the system driven by a homogeneous cw source.
The pump efficiency of such resonant bright solitons is estimated higher than that of the conventional DKS, at certain conditions of the loss modulation depths and the pump intensity.
We further employed the optical SA effect that would initiate a modulated loss profile along with the evolution of the system from the homogeneous state. In this way, we observed self-starting and stabilized multi-pulse pattern close to the Turing rolls at system's resonance, as well as other localized patterns.
In the red-detuned regime, soliton state close to the DKS was also characterized, which under the perturbation of the slow-SA effect would feature the energy exchange with adjacent soliton pulses.
Therefore, our work has demonstrated the role of the loss fluctuation in triggering the spontaneous symmetric breaking in dissipative and nonlinear systems, and initiating the formation of localized field patterns such as dissipative solitons.
Moreover, the self-starting soliton state operated on system's resonance would promise a decent conversion efficiency from a homogeneous cw pump to ultra-fast soliton pulses as well as to soliton micro-combs, which is highly desired for applications.
Yet, this work is limited to consider only the anomalous dispersion in the cavity.
The effect of transient loss fluctuation in normal dispersion cavities requires a further study.

From a prospective viewpoint, dissipative and nonlinear cavities in the presence of optical SA effect is accessible in the form of acitve semiconductor microcavities \cite{spinelli_spatial_1998}.
In addition, for dielectric microresonators, doping the cavity with active nanoparticles (by means of e.g. the atomic layer deposition, ALD \cite{george_atomic_2010, guha_surface-enhanced_2017}) may also introduce the SA effect to the system.
Indeed, the ALD process is not limited to active elements with a fast relaxation time, and is also open for high-gain and rear-earth elements in the slow relaxation mode \cite{kippenberg2006demonstration}.
Such samples are currently under preparation and will be introduced in a following work.
Yet, we are aware that the doping process as well as other surface processes may also change cavity properties including the intrinsic loss rate, the dispersion and the nonlinearity.
Therefore, in the current work, we only focus on a normalized model to investigate universal dynamics of stable localized structures that can be potentially accessed in future experiments.

\begin{backmatter}

\bmsection{Funding}
National Key Research and Development Program of China (2020YFA0309400); National Natural Science Foundation of China (11974234);
Shanghai Science and Technology Development Foundation (20QA1403500).


\bmsection{Disclosures}
The authors declare no conflicts of interest.

\bmsection{Data Availability Statement}
\textcolor{black}{Data and simulation codes related to this work are available in \cite{Dataset_2021}, or from the corresponding author upon reasonable request.}


\end{backmatter}




\end{document}